\documentclass[12pt]{article}

\usepackage{graphics}
\usepackage[dvips]{graphicx}

\def\sqr#1#2{{\vcenter{\hrule height.#2pt\hbox{\vrule width.#2pt height#1pt 
\kern#1pt 
\vrule width.#2pt}\hrule height.#2pt}}} 
\def\square{\mathchoice\sqr64\sqr64\sqr{4.2}3\sqr{3.0}3}

\begin{document} 
\title{The influence of scalar fields\\
in protogalactic interactions}  
\makeatletter	   
\author{M. A. Rodr\'\i guez-Meza\thanks{On leave from Instituto 
de F\'\i sica, Benem\'erita Universidad Aut\'onoma de Puebla.}, 
Jaime Klapp\thanks{e-mail:klapp@nuclear.inin.mx} and 
Jorge L. Cervantes-Cota\thanks{e-mail:jorge@nuclear.inin.mx}\\ 
Departamento de F\'{\i}sica, \\ 
Instituto Nacional de Investigaciones Nucleares (ININ) \\ 
Apartado postal 18--1027, M\'exico D.F. 11801, M\'exico \\ 
and \\
Heinz Dehnen \\ 
Fakult\"at f\"ur Physik \\ 
Universit\"at Konstanz, Postfach 5560 M 677 \\ 
D-78434 Konstanz, Germany } 
\date{\ }
\maketitle 
\makeatother     

\begin{abstract} 
We present simulations within the framework of scalar-tensor theories,
in the Newtonian limit, to investigate the influence of massive scalar
fields on the dynamics of the collision of two equal spherical clouds.
We employ a SPH code modified to include the scalar field
to simulate two initially non-rotating  
protogalaxies that approach each other, and as a 
result of the tidal interaction, intrinsic angular momentum is 
generated.  
We have obtained sufficient 
large values of $J/M$ to suggest that intrinsic angular momentum can 
be the result of tidal interactions. 
\end{abstract}

\section{Introduction} 
In recent years there has been achieved important progress in   
understanding the dynamics that led to the formation  
of galaxies. Two and three dimensional N-body simulations   
of galaxies and protogalaxies have been computed using up to a few 
millions of particles, giving a more realistic view of how galaxies,  
quasars, and black holes could have formed (Barnes \& Hernquist 1992, 
Barnes 1998).    
 
The Universe's composition at the time galaxies formed could be,  
theoretically, very varied, including baryonic visible and dark matter, non  
baryonic dark matter, neutrinos, and many cosmological relics  
stemming from symmetry breaking processes predicted by high energy  
physics (Kolb \& Turner 1990).   All these particles, if present, should  
have played a role in the structure formation.  Then, 
galaxies are expected to possess dark matter components and, in accordance   
with the rotational curves of stars and gas around the centres of spirals, they 
are in the form of halos (Ostriker \& Peebles 1973) and  must contribute 
to at least 3 to 10 times the mass of the visible matter (Kolb \& Turner 1990).    
 
Whatever the Universe composition was, protogalaxies were originated  
due to a spectrum of scale-invariant perturbations (Harrison 1970; 
Zel'dovich 1972) that was present within the  cosmological background  
at the beginning of structure formation; the inflationary cosmology is  
the most convincing scenario that explains its origin (Mukhanov et al 1992).    
Protogalactic structures began to acquire some momenta, e.g. tidal  
torques (Fall \& Efstathiou 1980), because of local gravitational  
instabilities to provoke plenty of collisions, mergings, fly-bys,  
etc, between these original, cosmic structures.  As a result of their evolution,   
galaxies, as we presently know them, must have formed.  The dynamics of  
protogalaxies has been studied intensively, for a review see   
Barnes \& Hernquist (1992). There has been much interest to understand how 
galaxies acquired their present features, especially how their  
internal angular momentum (spin) has been gotten,  
$J/M \sim {\cal{O}} (10^{30}) ~{\rm cm}^{2}/{\rm s}$.     
A very important issue about it is how the transfer of  
angular momentum between protogalaxies took place to give rise to the  
observed elliptical and spiral galaxies with their known mass and  
rotational properties.   As an initial condition can be thought that  
protogalaxies were gravitationally isolated. However, there are some  
indications that the orbital angular momentum in spiral galaxies in pairs  
is few times larger than their spins, so pairs seem to be not dynamically  
isolated (Oosterloo 1993).  Part of this angular momentum could had its origin  
in the cosmic expansion (Caimmi (1989,1990), Andriani \& Caimmi (1994)), where  
it has been computed the torques at the beginning of strong decoupling  
from the Hubble flow of spherical-symmetric density perturbations.  
Moreover, observations of spin angular momentum of various  
thousands of disc galaxies are compatible with the mechanism of  
generation of spin via tidal torques (Sugai \& Iye 1995).    
Other theoretical and numerical studies of evolution of angular momentum  
of protogalaxies from tidal torques are in line with observations  
(Chernin 1993, Catelan \& Theuns 1996). 

In the present work we investigate how the transfer of orbital to spin  
angular momentum is achieved when two equal, spherical clouds pass by,  
and in some cases when they collide; this type of interactions are to be  
expected in the tidal torques scheme.  
Studies of interacting spherical systems very related to ours 
have been done using the Newtonian theory of gravity, 
including a number of topics: mergings (White 1978, 1979),
mixing processes (White 1980), simulation of sinking satellites (White 1983a), and
mass and energy lost in tidal interactions (Aguilar \& White 1985), among others. 
However, we made our calculations within the framework of scalar-tensor 
theories, in the Newtonian limit, to investigate the influence  
of masive scalar fields on the dynamics.

This paper is organized as follows: In section 2 we present the 
Newtonian approximation of a typical scalar field theory.
In section 3, we present our models of protogalaxies,
the initial conditions, and the results.
The conclusions are in section 4. 

\section{Scalar Fields and the Newtonian Approximation}
We consider a typical scalar field theory given by the following Lagrangian
\begin{equation}
{\cal L} = \frac{\sqrt{-g}}{16\pi} \left[ -\phi R + \frac{\omega(\phi)}{\phi} 
	(\partial \phi)^2 - V(\phi) \right] + L_M(g_{\mu\nu})
\end{equation}
from which we get the gravity equations,
\begin{equation}
R_{\mu\nu} - \frac{1}{2} g_{\mu\nu} R = \frac{8\pi}{\phi}T_{\mu\nu} +
	\frac{V}{2\phi} g_{\mu\nu} + \frac{\omega}{\phi^2} \partial_\mu \phi \partial_\nu \phi
	-\frac{1}{2} \frac{\omega}{\phi^2}(\partial \phi)^2 g_{\mu\nu}
	+ \frac{\phi_{;\mu\nu}}{\phi} - \frac{g_{\mu\nu}}{\phi} \square \phi
\end{equation}
and the scalar field equation
\begin{equation}
\square \phi + \frac{\phi V' - 2V}{3+2\omega} = \frac{1}{3+2\omega} \left[
	8\pi T -\omega' (\partial \phi)^2 \right]
\end{equation}

We expect to have nowadays small deviations of the scalar fields around the background defined
here as $\langle \phi \rangle = 1$. If we define $\bar{\phi} = \phi -1$ the Newtonian approximation gives (Helbig 1991)
\begin{eqnarray}
R_{00} &=& \frac{1}{2} \nabla^2 h_{00} = 4\pi \rho - \frac{1}{2} \nabla^2 \bar{\phi} \\
\nabla^2 \bar{\phi} - m^2 \bar{\phi} &=& - 8\pi \alpha\rho
\end{eqnarray}
where we have done
\begin{displaymath}
\frac{\phi V' - 2V}{3+2\omega} = m^2 \bar{\phi} - m^2 k \bar{\phi}^2 + \ldots
\end{displaymath}
and $\alpha = 1 / (3 + 2\omega)$.

The solution of these equations is
\begin{eqnarray}
\bar{\phi} &=& 2 \alpha u_\lambda \nonumber \\
h_{00} &=& - 2u - 2\alpha u_\lambda
\end{eqnarray}
where
\begin{eqnarray}
u &=& \sum_a \frac{m_a}{| {\bf r} - {\bf r}_a |} \nonumber \\
u_\lambda &=& \sum_a \frac{m_a}{| {\bf r} - {\bf r}_a |} \exp \left[ -| {\bf r} - {\bf r}_a |/\lambda \right]
\end{eqnarray}
$\lambda = 1/m$, where $m$ is the mass given through the potential.
This mass can have a variety of values depending on the particular particle physics model.
The potential $u$ is the Newtonian part and $u_\lambda$ is the 
dark matter contribution which is of Yukawa type. 
The total force on a particle of mass $m_i$ is
\begin{equation}
\sum {\bf F} = -\frac{1}{2} \nabla h_{00} = m_i {\bf a} \quad .
\end{equation}
 
\section{Protogalactic Cloud Models and Results} 
Original protogalaxies could have   
very irregular forms, but we use as a first approximation spherical  
clouds for simplicity, and because this form seems to resemble the  
global shape of both visible and dark matter of many galaxies, i.e.,  
taking into account their spherical halos (Ostriker \& Peebles 1973, 
White 1983b).  The initial clouds are in polytropic equilibrium with 
small internal velocities, compared to what they would need to be in 
dynamical, gravitational equilibrium.  This feature avoids  
a large initial spin, in accordance with the fact that there were no  
primordial rotational motions in the universe (Ozernoy \& Chernin 1968;  
Parijskij 1973; Boynton \& Partridge 1973; Peebles \& Silk 1990).   
Then, the clouds are sent to approach each other, and only after their  
gravitational interaction takes place, spin will be gained.   

For the simulations, each protogalaxy is constructed using the Plummer model
given by the potential-density pair (Aarseth {\em et al} 1974),   
\begin{equation}
\Phi_P(r) = -\frac{GM}{\sqrt{r^2 + b^2}} \quad, \quad
\rho_P(r) = \left( \frac{3M}{4\pi b^3} \right) 
\left( 1 + \frac{r^2}{b^2} \right)^{-5/2}
\end{equation}
where $G$ is the gravitational constant, $M$ is the total mass and
$b$ is a parameter which determines the dimensions of the cloud.
Particle velocities are chosen everywhere isotropic which gives
a system initially in steady state. The total energy of the
cloud is $E=-(3\pi/64) G M^2 b^{-1}$.
We are using units in which $G=M=-4E=1$.  

We take two identical 3-D clouds consisting of $N=2^{11}$ particles.  
The initial separation between the clouds is  $10$, and the 
velocity of the center of masses are $V_1=(0.1,0.1,0)$ and 
$V_2=(-0.2,0.4,0)$ in our units. The initial velocities are 
given so that the kinetic energy is a fraction of the potential 
energy and we consider a range that is consistent with the 
observed velocities of galaxies in clusters that goes from 
50 km/s up to about 1000 km/s.  For the present investigation 
we consider that the protogalaxies moves initially in the plane 
$(x,y)$, and the angle of both protogalaxies is the same. More 
general initial conditions will be considered in a future 
communication.       
    
\begin{figure}
\includegraphics[width=5in]{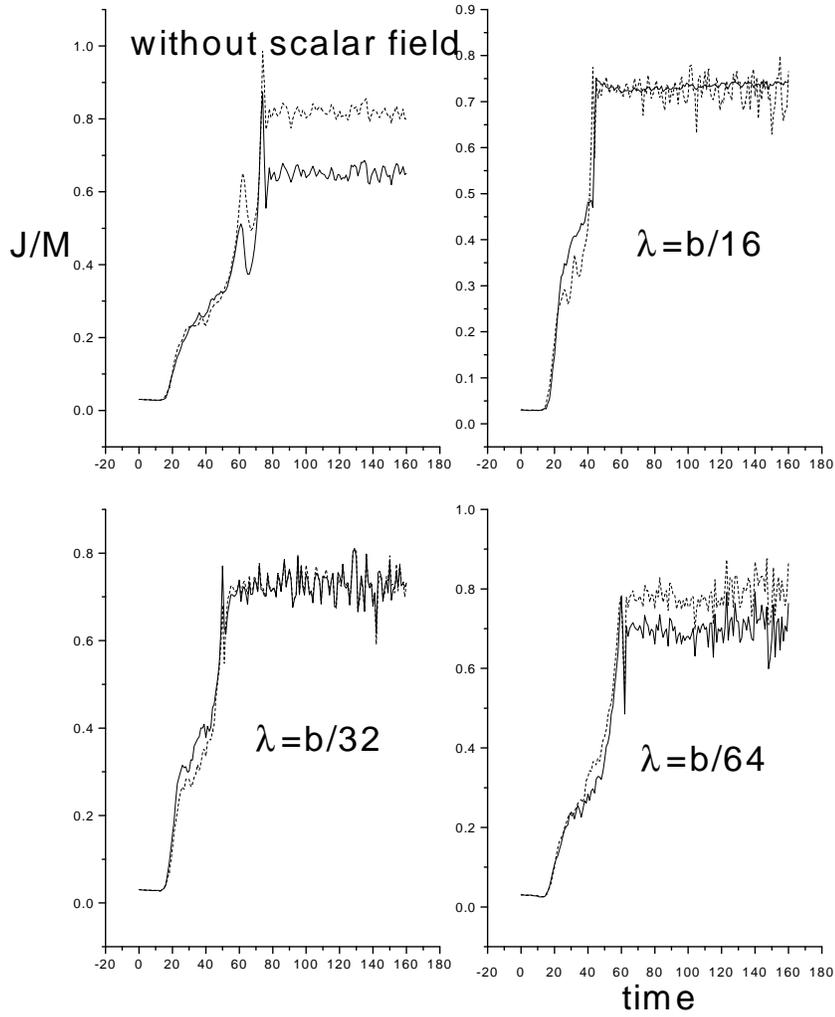}
\caption[Angular momentum]{Evolution of the angular momentum 
without
scalar field and with scalar field for different values
of $\lambda$}   
\label{fig1}
\end{figure}

A three-dimensional hydrodynamic code based on the TREESPH 
algorithm formulated by Hernquist \& Katz (1989) was used for 
the computations of this paper. The code combines the method of 
SPH, developed by Lucy (1977) and Gingold \& Monaghan (1977), 
with the hierarchical tree algorithm of Hernquist (1987) for 
the calculation of the gravitational acceleration forces. The 
SPH method is a grid-free Lagrangian scheme in which the physical 
properties of the fluid are determined from the properties of a 
finite number of particles. In order to reduce the statistical 
fluctuations resulting from representing a fluid continuum by a 
finite set of particles, a smoothing procedure is employed in 
which the mean value of a field quantity is estimated from its 
local values by kernel interpolation. Thus, the evolution of 
particle $i$ is determined by solving Euler's equation 
\begin{eqnarray} 
\frac{d{\bf r}_{i}}{dt}&\equiv&{\bf v}_{i}\hspace{0.2 cm}\\ 
\frac{d{\bf v}_{i}}{dt}&=&-\frac{1}{\rho}\nabla p_{i}-
\frac{1}{2}\nabla (h_{00})_i + 
{\bf A}_{visc,i}\hspace{0.2 cm}, 
\end{eqnarray} 
where $p_{i}$ and ${\bf A}_{visc,i}$ denote, respectively, the 
gas pressure and the artificial viscous acceleration associated 
with particle $i$. This quantities are introduced because we are 
considering that the protogalaxies are gaseous. The code was modified 
(Rodr\'\i guez-Meza 2000)
to include the effect of the scalar field, through the term $h_{00}$
given by Eq.\ (6). 
 
The simulation of the interaction of two protogalactic models starts 
with the clouds separated by a distance of 10 and on the $x$-axis. 
The selected separation is large 
enough to ensure that tidal effects are important but small enough 
that the calculation is possible in a  reasonable computing 
time.  Each particle in the initial steady state clouds  
is given an additional velocity ($V_{1}$ or $V_{2}$, corresponding to  
cloud 1 or 2) so that its magnitude is  
much bigger than the internal velocities they have at the  
equilibrium described above.  In this way, initial 
clouds are almost spinless, and the given  velocities $V_{1}, ~V_{2}$ 
imply kinetic energies associated with each cloud.  The evolution of the
intrinsic angular 
momentum $(J/M)$ with  respect to the center of mass of each cloud is shown
in Fig.\ 1. Continuous lines indicate cloud 1 and dashed lines cloud 2. 
The first plot is without scalar field, the other plots 
consider values of $\lambda$ of $b/16$, $b/32$, and $b/64$, respectively.
We observe that the intrinsic angular momenta start from their initial
values to a constant mean value approximately of 0.75 which in the cgs units
is of the order of $10^{30}$ cm$^2$/s.
This transient stage is slower without
scalar field than the ones which consider scalar field. The faster transient
occurs when $\lambda$ is bigger. 
In Fig.\ 2 we show plots of phase space $v_r$ versus $r$ of the 
whole system and for the same cases as Fig.\ 1. The scalar field 
extends the phase space in the $v_r$ direction.

\begin{figure}
\includegraphics[width=5in]{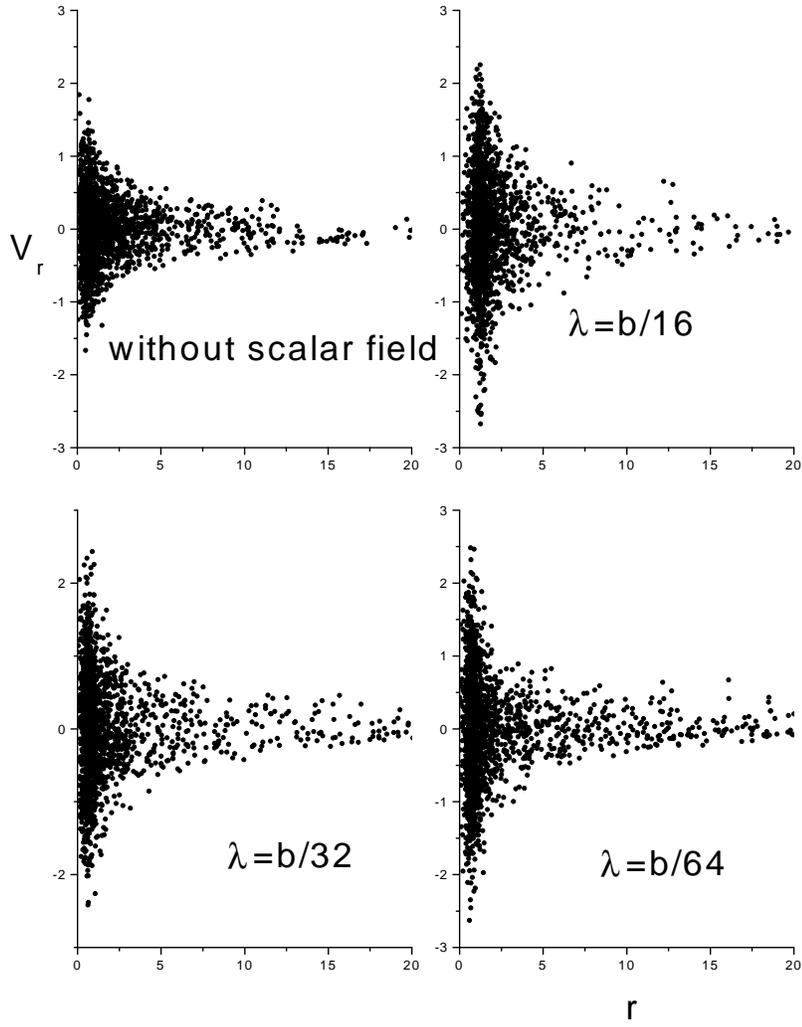}
\caption[Angular momenutm]{Phase space $v_r$ versus $r$ of the combined 
system after the transient stage without scalar field and with scalar 
field for different values
of $\lambda$}   
\label{fig2}
\end{figure}

\section{Conclusions}
We have made other computations varying the kinetic  
energy from 4 to 1/16 times the potential energy.  For  
large values of the kinetic energy the deflection is small, but for  
small values there is a considerable deflection, and in some cases  
we got almost a head-on collision. This is consistent with the known
fact that the merging probability in an encounter of two clouds is enhanced
significantly when the encounter takes place at relatively low speed (see
for instance Makino \& Hut 1997).
We found that  
only close encounters and  mergings permit the original spinless clouds to 
gain rotational velocities as is observed in typical galaxies nowadays.   
Similar studies have being done (Namboodiri \& Kochhar 1991) considering 
point mass perturbers.  In our approach the perturber is itself a 
protogalaxy and therefore the dynamics is more complicate, especially in close 
encounters.  
We have also found that the transient time to spin up the clouds depends
on the scalar field. The transient stage is
faster than the one without scalar field. When the scalar field is included
faster transients occur for bigger values of $\lambda$. The phase space
$v_r$ versus $r$ of the combined system is also more extended in the 
$v_r$ direction with scalar field than the one without scalar field.
 
\vspace {1 cm} 
    
{\bf Acknowledgments} 
\bigskip 
    
The computations of this paper were performed using the Silicon 
Graphics Oring2000 computer of the Instituto de F\'\i sica,
Benem\'erita Universidad Aut\'onoma de Puebla, M\'exico. 
This work was supported in part by the DAAD and CONACyT grant 
numbers 33278-E and 33290-E.


\end{document}